\documentclass[11pt]{article}
\usepackage{amsmath,amssymb,color,graphics,epsfig,cite}

\def\baselinestretch{1.4}
\usepackage{amsfonts}

\newcommand{\hoch}[1]{$\, ^{#1}$}

\makeatletter
\@addtoreset{equation}{section}
\makeatother

\newcommand{\be}{\begin{equation}}
\newcommand{\ee}{\end{equation}}
\newcommand{\bea}{\setlength\arraycolsep{2pt} \begin{eqnarray}}
\newcommand{\eea}{\end{eqnarray}}

\def\ft#1#2{{\textstyle{\frac{\scriptstyle #1}{\scriptstyle #2} } }}
\def\fft#1#2{{\frac{#1}{#2}}}

\def\0{{\sst{(0)}}}
\def\1{{\sst{(1)}}}
\def\2{{\sst{(2)}}}
\def\3{{\sst{(3)}}}
\def\4{{\sst{(4)}}}
\def\5{{\sst{(5)}}}
\def\6{{\sst{(6)}}}
\def\7{{\sst{(7)}}}
\def\8{{\sst{(8)}}}
\def\sst#1{{\scriptscriptstyle #1}}

\begin{document}

\begin{flushright}
\hfill{}

\end{flushright}

\begin{center}
{\Large {\bf Dyonic (A)dS Black Holes in Einstein-Born-Infeld Theory in 
		Diverse Dimensions}}

\vspace{15pt}
{\bf Shoulong Li\hoch{1}, H. L\"u\hoch{2, \ast} and Hao Wei\hoch{1}}

\vspace{10pt}

\hoch{1}{\it School of Physics, \\
Beijing Institute of Technology, 5 South Zhongguancun Street, Beijing 100081, China}

\vspace{10pt}

\hoch{2}{\it Center for Advanced Quantum Studies, Department of Physics, \\
Beijing Normal University, 19 Xinjiekouwai Street, Beijing 100875, China}

\vspace{20pt}

\underline{ABSTRACT}

\end{center}

We study Einstein-Born-Infeld gravity and construct the dyonic (A)dS 
planar black holes in general even dimensions, that carry both the 
electric charge and magnetic fluxes along the planar space. In four 
dimensions, the solution can be constructed with also spherical and
 hyperbolic topologies.  We study the black hole thermodynamics and 
 obtain the first law.  We also classify the singularity structure.

\vfill  {\footnotesize sllee\_phys@bit.edu.cn \ \ 
	\ \hoch{\ast}mrhonglu@gmail.com \ \ \ haowei@bit.edu.cn}

\thispagestyle{empty}

\pagebreak

\tableofcontents
\addtocontents{toc}{\protect\setcounter{tocdepth}{2}}



\section{Introduction}

In 1934, Born and Infeld~\cite{Born:1934gh} proposed an elegant 
nonlinear version of electrodynamics that successfully removes 
the divergence of self-energy of a point-like charge in Maxwell's
 theory of electrodynamics. The Lagrangian density of the 
 Born-Infeld (BI) theory in $D$-dimensional Minkowski spacetime
  is given by
\be
{\cal L} = -b^2\sqrt{-\det\left(\eta_{\mu\nu} + 
	\frac{F_{\mu\nu}}{b}\right)} +b^2 \,, \label{eq1}
\ee
where $\eta_{\mu\nu} = \textup{diag}(-1, 1, 1, 1)$ is the 
Minkowski metric, $F_{\mu\nu}= 2 \partial_{[\mu} A_{\nu]}$
 is the Faraday tensor and $A=A_\mu dx^\mu$ is the Maxwell
  gauge potential.  BI theory contains a dimensionful parameter
   $b$, and in the limit $b\rightarrow \infty$, BI theory reduces
    to the Maxwell theory,
\be
{\cal L} = -\frac{1}{4} F^2 +{\cal O}\left(\frac{1}{b^2}\right) \,.\label{eq2}
\ee
In the limit $b\rightarrow 0$, the Lagrangian in four dimensions
 becomes $F\wedge F$ which is a total derivative.  The limit is
  generally singular in higher dimensions.

BI theory has enjoyed further attentions since the invention of
 string theory. It turns out that the BI action can arise from 
 string theory~\cite{Fradkin:1985qd}, describing the low energy 
 dynamics of D-branes~\cite{Leigh:1989jq}. We refer to 
 {\it e.g.}~\cite{Gibbons:2001gy, Tseytlin:1999dj} for some 
 comprehensive reviews on the BI theory in string theory.  The 
 special Born-Infeld-like nonlinear form is also very useful to
  construct analogous new theories, such as Dirac-Born-Infeld~(DBI)
   inflation theory~\cite{Silverstein:2003hf, Alishahiha:2004eh} 
   and  Eddington-inspired Born-Infeld~(EiBI) 
   cosmologies~\cite{Banados:2010ix}. BI theory can also be adopted
    to explore issues of dark energy~\cite{Elizalde:2003ku, Fuzfa:2006pn}.

In this paper, we focus on the study of black holes in 
Einstein-Born-Infeld~(EBI) theory. The most general static type-$D$ 
metric of the BI theory in four dimensions was constructed in~\cite{GSP1984}. 
(See also~\cite{Cataldo:1999wr} and~\cite{Fernando:2003tz}.) The 
spherically-symmetric solution was generalized to arbitrary $D$ dimensions 
in~\cite{Dey:2004yt} where the black hole thermodynamics was studied. The 
black hole solutions was also generalized to include different 
topologies~\cite{Cai:2004eh}. The Born-Infeld black hole solutions were also 
studied in Einstein theory with a dilaton field~\cite{Dehghani:2006zi} and
 in the modified gravity theories such as Gauss-Bonnet theory~\cite{Dehghani:2006ke},
 Lovelock theory~\cite{Dehghani:2008qr}, Brans-Dicke theory~\cite{Hendi:2015hgg},
 $f(T)$ theory~\cite{Junior:2015dga}, massive gravity~\cite{Hendi:2015hoa}, and so on. The extended 
 thermodynamics~\cite{Dey:2004yt, Cai:2004eh, Dehghani:2006zi, Dehghani:2006ke,
 Dehghani:2008qr, Hendi:2015hgg, Junior:2015dga, Zou:2013owa, 
 Gunasekaran:2012dq, Banerjee:2011cz}, geodesics~\cite{Breton:2002td, Linares:2014nda},
 and AdS/CFT correspondence properties~\cite{Cai:2008in, Jing:2010zp, 
 	Chaturvedi:2015hra} were studied too. Other Born-Infeld 
 solutions are also studied, for  example, thin-shell wormholes~\cite{Eiroa:2011aq}.

(A)dS black hole solutions in BI theory considered in literature typically 
involves only either the electric or magnetic charges. Although the dyonic 
black hole in EBI theory was constructed in~\cite{GSP1984}, it is written 
in the general (static) type-$D$ form. The global structure in the 
spherically symmetric form was analysed in~\cite{Breton:2002td} for 
the asymptotically-flat case. In this paper, we shall first study the 
dyonic (A)dS black holes in the EBI theory in four dimensions with 
general topologies, focus on analysing the black hole thermodynamics 
and singularity structure. We then construct dyonic AdS planar black 
holes in arbitrary even dimensions, where the solutions carry both the 
electric flux as well as the magnetic 2-form flux along the planar space.

Interestingly in almost all the previous works on constructing black holes, 
the equivalent action in four dimensions was used, rather than the original
 one. In $D=4$, the Lagrangian can be equivalently expressed as~\cite{Born:1934gh}
\be
{\cal L} =   b^2 -b^2\sqrt{1 + I_1 + I_2} \,\,,\label{eq3}
\ee
where
\begin{equation}
I_1 = \frac{1}{2b^2} F_{\mu\nu} F^{\mu\nu} = \frac{B^2 - 
	E^2}{b^2} \,,\qquad
I_2 = -\frac{1}{16 b^4} \left(F_{\mu\nu} \widetilde{F}^{\mu\nu} 
\right)^2 = -\frac{\left(E \cdot B \right)^2}{b^4} \,,\label{eq4}
\end{equation}
in which $E$ and $B$ are electric and magnetic fields, and
\be
\widetilde{F}^{\mu\nu} = \frac{1}{2} \epsilon^{\mu\nu\rho\sigma} 
F_{\rho\sigma} = \frac{1}{2\sqrt{-\det(\eta_{ab})}} 
\varepsilon^{\mu\nu\rho\sigma} F_{\rho\sigma}\,,\label{eq5}
\ee
where $\varepsilon^{\mu\nu\rho\sigma}$ is a tensor density with
 $\varepsilon^{0123} = 1$.

The equivalence of (\ref{eq3}) and (\ref{eq1}) is only true in four 
dimensions; it is no longer valid in higher dimensions.  However, if
 one considers only static solutions carrying electric charges, one 
 can nevertheless use the reduced Lagrangian (\ref{eq3}). In fact in 
 this case, one can even ignore the $I_2$ term.  This was indeed done 
 in many previous works, for example~\cite{Cataldo:1999wr, 
 Fernando:2003tz, Dey:2004yt, Cai:2004eh}.  Since one of our purposes 
is to construct dyonic black holes in higher dimensions, the Lagrangian 
(\ref{eq3}) is not suitable for this purpose and we shall use the original 
Lagrangian (\ref{eq1}) instead for all our constructions.

The paper is organized as follows. In Sec.~\ref{sec2}, we review the EBI 
theory and then derive the equations of motion for all dimensions. In 
Sec.~\ref{sec3}, we obtain the exact dyonic (A)dS black hole solutions in 
four dimensions with a generic topological horizon. Then we study the global 
structure, black hole thermodynamics and the singularity structures. In 
Sec.~\ref{sec4}, we generalize the results to all even dimensions. We conclude
 the paper in Sec.~\ref{sec5}.


\section{EBI and its equations of motion}\label{sec2}

In this section, we consider the EBI theory. The Lagrangian of BI theory can 
be naturally generalized to curved spacetimes and the Lagrangian is given by
\be{}
{\cal L} = -b^2\sqrt{-\det\left(g_{\mu\nu} + \frac{F_{\mu\nu}}{b}\right)} +b^2 
\sqrt{-\det \left(g_{\mu\nu}\right)} \,\,,\label{eq6}
\ee
where $g_{\mu\nu}$ is the metric. The Lagrangian of the EBI theory with a bare 
cosmological constant $\Lambda_0$  can be written by
\be{}
{\cal L} =   \sqrt{-g}\left(R - 2\Lambda_0\right) -b^2\sqrt{-\det\left(g_{\mu\nu} 
	+ \frac{F_{\mu\nu}}{b}\right)} \,,\label{eq7}
\ee
where $\Lambda_0=\Lambda-b^2/2$.  Here, $\Lambda$ is the effective cosmological 
constant.
The variation of Lagrangian (\ref{eq7}) gives rise to
\begin{equation}
\delta {\cal L} = \sqrt{-g}\left(-E^{\mu\nu} \delta g_{\mu\nu} +E_A^\nu \delta 
A_\nu + \nabla_\mu J^\mu \right) \,,\label{eq8}
\end{equation}
where $g = \det(g_{\mu\nu})$, $J^\mu$ is the surface term and
\begin{align}
E^{\mu\nu} &= G^{\mu\nu} +g^{\mu\nu} \Lambda_0 +\frac{b^2}{2} \frac{\sqrt{-h}}
{\sqrt{-g}} \,\left(h^{-1} \right)^{(\mu\nu)} \,, \label{eq9}\\
E_A^\nu &= \nabla_\mu \left[\frac{\sqrt{-h}}{\sqrt{-g}}\, b \left(h^{-1} 
\right)^{[\mu\nu]}\right] \,, \label{eq10}
\end{align}
in which $G^{\mu\nu} = R^{\mu\nu} - g^{\mu\nu} R / 2$, $h_{\mu\nu} = g_{\mu\nu} 
+F_{\mu\nu}/b$, $h \equiv \det(h_{\mu\nu})$, and $(h^{-1})^{\mu\nu}$ denotes the
 inverse of $h_{\mu\nu}$, satisfying
\be
(h^{-1})^{\mu\rho}\, h_{\rho\nu}=\delta^\mu_\nu\,,\qquad
h_{\nu\rho}\, (h^{-1})^{\rho\mu}=\delta^\mu_\nu\,.
\ee
We further defined
\be{}
\left(h^{-1} \right)^{(\mu\nu)} = \frac{1}{2}\left[ \left(h^{-1} \right)^{\mu\nu}
 +\left(h^{-1} \right)^{\nu\mu}\right] \,,\qquad \left(h^{-1} \right)^{[\mu\nu]}
  = \frac{1}{2}\left[ \left(h^{-1} \right)^{\mu\nu} -\left(h^{-1} 
  \right)^{\nu\mu}\right] \,.\label{eq11}
\ee
The equations of motion are then given by $E^{\mu\nu}=0$ and $E_A^\mu=0$. These
 equations are derived from the original Lagrangian (\ref{eq7}) of the EBI theory 
 and hence are applicable in all dimensions and for all charge configurations.


\section{Dyonic black hole in four dimensions}\label{sec3}

In the previous section, we obtained the equations of motion of the EBI theory. We
 now construct the static dyonic (A)dS black hole solution with a general 
 topological horizon in four dimensions. We shall then study the global structure 
 and the black hole thermodynamics.


\subsection{Local solution}\label{sec3A}

The static solution in the type-$D$ form in the EBI theory was first constructed 
in \cite{GSP1984}.
The spherically-symmetric and asymptotically-flat solution was given 
in~\cite{Breton:2002td}.  In this section, we study the properties of the dyonic 
(A)dS black holes. The most general static ansatz can be written as
\be{}
ds^2 = -h(r) dt^2 +\frac{dr^2}{f(r)} +r^2\left(\frac{du^2}{1 - k u^2} +(1 - 
k u^2) d \varphi^2\right), \qquad A = \phi(r) dt +p u d\varphi \,, \label{eq12}
\ee
where $k = 1, 0, -1$ denotes the metric for the unit 2-spheres, 2-torus or the 
unit hyperbolic 2-space, and $p$ is magnetic charge parameter. It turns out that 
the equations of motion of the metric $g_{\mu\nu}$ imply that $h(r) = f(r)$ and 
the equations of motion for $A_{\mu}$ imply that $\phi(r)$ can be expressed as
\be{}
\phi^\prime(r) = \frac{q}{\sqrt{r^4 +\frac{Q^2}{b^2}}} \,,\qquad \textup{with}
 \qquad Q = \sqrt{p^2 + q^2}\,, \label{eq13}
\ee
where and thereafter, we use a prime to denote a derivative with respect to $r$, 
and $q$ is a integral constant that is related to the electric charge. The 
function $f(r)$ satisfies
\be{}
r f^\prime(r) + f(r) = k  - \Lambda_0 r^2 - \frac{b^2}{2}\sqrt{r^4 
	+\frac{Q^2}{b^2}} \,. \label{eq14}
\ee
Thus $f(r)$ can be solved and expressed in terms of hypergeometric function ${}_2 F_1$
\begin{equation}
 f(r) = -\frac{1}{3}\Lambda_0 r^2 + k - \frac{\mu}{r} - \frac{b^2}{6} \sqrt{r^4 
 	+\frac{Q^2}{b^2}} +\frac{Q^2}{3 r^2} \, {}_2 F_1 \left[\frac{1}{4}, 
 \frac{1}{2}; \frac{5}{4}; -\frac{Q^2}{b^2 r^4}\right] \,, \label{eq15}
\end{equation}
where $\mu$ is the integral constant corresponding to the  mass of the 
solution. The electric potential $\phi(r)$ is expressed by
\be{}
\phi(r) = \int_{r}^{\infty}\frac{q d\tilde{r}}{\sqrt{\tilde{r}^4 
+\frac{Q^2}{b^2}}} = \frac{q}{r}\, {}_2 F_1 \left[\frac{1}{4}, \frac{1}{2};
 \frac{5}{4}; -\frac{Q^2}{b^2 r^4}\right] \,. \label{eq16}
\ee
In the limit $b\rightarrow\infty$, the solution recovers the dyonic 
Reissner-Nordstr\"om-(A)dS black hole,
\be{}
f(r) = -\frac{\Lambda}{3} r^2 + k - \frac{\mu}{r} + \frac{Q^2}{4 r^2} 
\,,\qquad \, \phi(r) = \frac{q}{r} \,. \label{eq17}
\ee
On the other hand, in the limit $b\rightarrow0$, the Born-Infeld field 
vanishes and the solution is reduced to the Schwarzschild-(A)dS black 
hole in pure cosmological Einstein theory,
\be{}
f(r) = k - \frac{\mu}{r} - \frac{\Lambda}{3}r^2 \,.\label{eq18}
\ee
In the large-$r$ expansion, we have
\be{}
f(r) = -\frac{\Lambda}{3} r^2 + k - \frac{\mu}{r} + \frac{Q^2}{4 r^2} 
+{\cal O}\left(\frac{1}{r^6}\right)\,,\qquad \, \phi(r) = \frac{q}{r}  
+{\cal O}\left(\frac{1}{r^5}\right)\,.\label{eq19}
\ee
Thus we see that the first few leading-order expansions match those of 
the Reissner-Nordstr\"om-(A)dS black hole.

\subsection{Thermodynamics}\label{sec3B}

Now we discuss black hole thermodynamics. The event horizon 
is defined through $f(r_+) = 0$, where $r_+$ denotes the largest root of 
$f$.  It is convenient to express the constant $\mu$ in terms of $r_+$, 
namely
\be{}
\mu = -\frac{1}{3}\Lambda_0 r_+^3 + k r_+  - \frac{b^2 r_+ }{6} \sqrt{r_+^4 
	+\frac{Q^2}{b^2}} +\frac{Q^2}{3 r_+} \, {}_2 F_1 \left[\frac{1}{4}, 
\frac{1}{2}; \frac{5}{4}; -\frac{Q^2}{b^2 r_+^4}\right] \,.\label{eq20}
\ee
Since the metric is asymptotically (A)dS, according to the definition of mass 
in asymptotically (A)dS space by Abbott-Deser-Tekin (ADT) 
formalism~\cite{Abbott:1981ff}, we find
\be{}
M = \frac{\omega_2}{8\pi}\mu \,,\label{eq21}
\ee
where $\omega_2 = \int du d\varphi$.  For $k=1$, corresponding the unit 
$S^2$, we have $\omega_2=4\pi$.

The temperature $T$ and entropy $S$ on the horizon are easily calculated as
 \begin{align}
 T &= \frac{f^\prime(r_+)}{4\pi} =\frac{k - \Lambda_0 r_+^2}{4\pi r_+} 
 -\frac{b^2}{8\pi r_+} \sqrt{r_+^4 + \frac{Q^2}{b^2}} \,, \label{eq22} \\
S &= \frac{{\cal A}}{4} = \frac{r_+^2}{4} \omega_2 \,.\label{eq23}
\end{align}
 The electric and magnetic charges are given by
\be{}
Q_e = \frac{\omega_2}{16\pi}\sqrt{-h}(h^{-1})^{[tr]}|_{r\rightarrow\infty} = 
\frac{q}{16\pi}\omega_2 \,, \qquad Q_m = \frac{\omega_2}{16\pi}F_{u\phi}|_{r\rightarrow\infty}
 = \frac{p}{16\pi}\omega_2 \,.\label{eq24}
\ee
Note that the above electric charge as a conserved quantity follows from the 
equation of motion (\ref{eq10}).  The electric and magnetic potentials are 
given by
\begin{equation}
\Phi_e = \frac{q}{r_+}\, {}_2 F_1 \left[\frac{1}{4}, \frac{1}{2}; \frac{5}{4}; 
-\frac{Q^2}{b^2 r_+^4}\right] \,, \qquad
\Phi_m = \frac{p}{r_+}\, {}_2 F_1 \left[\frac{1}{4}, \frac{1}{2}; \frac{5}{4};
 -\frac{Q^2}{b^2 r_+^4}\right] \,.
\end{equation}
The differential first law of black hole thermodynamics can be written as
\be
d M = T d S + \Phi_e d Q_e + \Phi_m d Q_m \,.\label{eq27}
\ee
One can further treat the cosmological constant as a generalized ``pressure" 
${\cal P}_{\Lambda _0}= -\Lambda_0/(8\pi)$ \cite{Cvetic:2010jb, Kastor:2009wy}. 
The conjugate quantity ${\cal V}$ can be viewed as a thermodynamical volume. 
The first law reads
\be{}
d M = T d S + \Phi_e d Q_e + \Phi_m d Q_m +{\cal V} d 
{\cal P}_{\Lambda_0} \,,\label{eq28}
\ee
where
\be{}
{\cal V} = \frac{\omega_2}{3}r_+^3 \,.\label{eq29}
\ee
Since $b$ is a dimensionful quantity, it will inevitably appearing in the 
Smarr relation.
It is useful also to introduce it as a thermodynamical quantity. Since $b^2$ 
has the same dimension of the cosmological constant, we may define 
${\cal P}_{b}=- b^2/(16\pi)$,
The corresponding thermodynamical potential is
\be{}
 {\cal V}_b=  \frac{\omega_2}{3}r_+^3\left(\sqrt{1 + \frac{Q^2}{r_+^4 b^2}} 
 -\frac{Q^2}{2 b^2 r_+^4}\, {}_2 F_1 \left[\frac{1}{4}, \frac{1}{2}; 
 \frac{5}{4}; -\frac{Q^2}{b^2 r_+^4}\right]
\right)\,.\label{eq30}
\ee
The extended differential first law of black hole thermodynamics is given by
\be{}
d M = T d S + \Phi_e d Q_e + \Phi_m d Q_m +{\cal V} d {\cal P}_{\Lambda_0} 
+{\cal V}_b d{\cal P}_b\,.\label{eq31}
\ee
The above first law can also be expressed as
\be{}
d M = T d S + \Phi_e d Q_e + \Phi_m d Q_m +{\cal V} d {\cal P}_{\Lambda} 
+\fft{b}{8\pi}({\cal V}-{\cal V}_b) db\,,
\ee
as was proposed in \cite{Gunasekaran:2012dq}.  The integral first law of black 
hole thermodynamics , also called Smarr formula, is given by
\be{}
M = 2\left(T S - {\cal V} {\cal P}_{\Lambda_0} - {\cal V}_b\, {\cal P}_b\right) 
+ \Phi_e Q_e + \Phi_m Q_m  \,.\label{eq32}
\ee
(See, also \cite{Rasheed:1997ns, Breton:2004qa, YiHuan:2010zz}.) When the 
topological parameter $k=0$, corresponding to AdS planar black holes, there 
exists an additional generalized Smarr relation \cite{Liu:2015tqa}
\be
M=\fft23 (T S + \Phi_e Q_e + \Phi_m Q_m)\,. \label{eq32a}
\ee


\subsection{Wald formalism}\label{sec3C}

Now we will calculate the conserved charge $M$ by using Wald 
formalism~\cite{Abbott:1981ff}. The conserved charges of AdS black hole has 
been calculated by many different methods such as the covariant phase space 
approach~\cite{Wald:1993nt, Iyer:1994ys} developed by Wald, ADT 
formalism~\cite{Abbott:1981ff}, and quasi-local ADT 
formalism~\cite{Kim:2013zha, Kim:2013cor, Peng:2014gha, Wu:2015ska, Peng:2015yjx, Peng:2016wzr}.
 The Wald formalism has been used to study the first law of thermodynamics for 
 asymptotically-AdS formalism in lots of theories, including Einstein-scalar 
 theoy~\cite{Liu:2013gja, Lu:2014maa}, Einstein-Proca~\cite{Liu:2014tra}, 
 Einstein-Yang-Mills~\cite{Fan:2014ixa}, 
 Einstein-Horndeski~\cite{Feng:2015oea, Feng:2015wvb, Feng:2015sbw}, in gravities
  extended with quadratic-curvature invariants~\cite{Fan:2014ala}, and also for 
  Lifshitz black hole~\cite{Liu:2014dva}.

Since the conserved charge of dyonic black hole has the same as that with pure 
electric case, so for simplicity we calculate the conserved charge for the back 
hole with the pure electric charge.  The effective Lagrangian is 
\be{}
{\cal L} = \sqrt{-g} L \,, \qquad L = R - 2\Lambda_0  -b^2\sqrt{1 + 
	\frac{F^2}{2b^2}} \,,\label{eq32b}
\ee
A general variation of the Lagrangian (\ref{eq32b}) was given in (\ref{eq8}). 
The equations of motion are given by
\begin{align}
E_{\mu\nu} &= G_{\mu\nu} +g_{\mu\nu}\Lambda_0 +\frac{1}{2}g_{\mu\nu}b^2\sqrt{1 
	+ \frac{F^2}{2b^2}} -\frac{F_{\mu\rho}{F_\mu}^{\rho}}
{2\sqrt{1+\frac{F^2}{2b^2}}} \,, \\
E_A^\nu &= \nabla_\mu\left( \frac{F^{\mu\nu}}{\sqrt{1+\frac{F^2}{2b^2}}} \right) \,.
\end{align}
The surface term $J^\mu = J_g^\mu + J_A^\mu$ is given by
\begin{equation}\begin{split}
J_g^\mu &= g^{\mu\rho}g^{\nu\sigma}\left(\nabla_\sigma \delta g_{\nu\rho} - 
\nabla_\rho \delta g_{\nu\sigma}\right) \,,\\
J_A^\mu &= -\frac{ F^{\mu\nu} \delta A_\nu}{\sqrt{1+\frac{F^2}{2b^2}}}  \,.
\end{split}\end{equation}
From this one can define a 1-form $J_{(1)} = J_\mu d x^\mu$ and its Hodge dual 
$\Theta_{(D-1)} = (-1)^{(D-1)}\star J_{(1)}$. Considering the infinitesimal 
diffeomorphism $x^\mu\rightarrow x^\mu + \xi^\mu$, one can get
\be{}
J_{(D-1)} \equiv \Theta_{(D-1)} - i_\xi \star {\cal L} = E_\Phi \delta \Phi - 
d\star J_{(2)} \,,
\ee
where  $i_\xi$ denotes a contraction of $\xi^\mu$ on the first index of the 
$D$-form $\star {\cal L}$. One can thus define an $(D-2)$-form ${\cal Q}_{(D-2)} = 
\star J_{2}$ and $J_{(D-1)} = d{\cal Q}_{D-2}$. Here we use the subscript notation 
``${}_{(p)}$" to denote a $p$-form. To make contact with the first law of 
black hole thermodynamics, we take $\xi^\mu = (\partial_t)^\mu$. Wald shows 
that the variation of the Hamiltonian with respect to the integration constants 
of a specific solution is given by
\be{}
\delta {\cal H} = \frac{1}{16\pi} \delta \int_c J_{(D-1)} - \frac{1}{16\pi} 
\int_c d\left( i_\xi \Theta_{(D-1)}\right) = \frac{1}{16\pi} \int_{\Sigma^{(D-2)}} 
\left(\delta {\cal Q}_{(D-2)} -i_\xi \Theta_{(D-1)} \right) \,,
\ee
where $c$ denotes a Cauchy surface and $\Sigma^{(D-2)}$ is its boundary, which 
has two components, one at infinity and one on the horizon. Thus according to 
the Wald formalism, the first law of black hole thermodynamics is a consequence of
\be{}
\delta {\cal H}_\infty = \delta {\cal H}_+ \,.
\ee
For four dimensional EBI theory, we have
\be{}
J_{\alpha_1\alpha_2\alpha_3} = \text{E.O.M.} + \epsilon_{\alpha_1\alpha_2\alpha_3\mu}
\nabla_\nu \left(2 \nabla^{[\nu}\xi^{\mu]} - 
\frac{F^{\mu\nu}A_\lambda\xi^\lambda}{\sqrt{1+\frac{F^2}{2b^2}}} \right) \,.
\ee
To specialise to our static black hole ansatz (\ref{eq12}) in $D=4$ dimensions 
(note that $h(r) = f(r)$), the result for Lagrangian is well established and is 
given by
\be{}
\delta {\cal Q} - i_\xi \Theta = -\omega_2 r^2 \left(\frac{2\delta f}{r} +\left[1- 
\frac{{\phi^\prime}^2}{b^2} \right]^{-\frac{3}{2}} r^2 \phi \delta \phi^\prime 
\right) \,.
\ee
Choosing the gauge such that the electrostatic potential $\phi$ vanishes on 
the horizon, it is straightforward to verify that
\be{}
\delta {\cal H}_+ = T \delta S\,,\qquad \delta {\cal H}_\infty = \delta M- 
\Phi_e \delta Q_e \,,
\ee
which yields the first law of black hole thermodynamics $dM=TdS + \Phi_e dQ_e$.

\subsection{Singularity structures}

Although the vector field is singularity free, the general solution has a 
curvature singularity at the origin $r=0$.  To study the nature of the singularity, 
we consider small-$r$ expansion near the origin:
\be
f=-\fft{2(M-M^*)}{r} + k - \ft12 b Q + \ft16 (b^2 - 2\Lambda_0) r^2 - 
\fft{b^3}{20Q} r^4 +
{\cal O}(r^8)\,,
\ee
where
\be
M^* = \fft{\Gamma(\fft14)^2\,\sqrt{b}}{24\sqrt{\pi}}\,  Q^{\fft32}\,.
\ee
The Riemann-tensor squared is given by
\be
R^{\mu\nu\rho\sigma}R_{\mu\nu\rho\sigma} =  \fft{48(M-M^*)^2}{r^6} + 
\fft{8bQ(M-M^*)}{r^5} +
\fft{b^2 Q^2}{r^4} + {\cal O}(\fft{1}{r^2})\,.
\ee
Thus we see that when $M> M^*$, the spacetime has a space-like singularity 
analogous to the Schwarzschild black hole, whilst it has a time-like 
singularity.  Note that the time-like singularity arising from $M< M^*$ is 
different from that in the Reissner-Nordstr\"om black hole which has a $1/r^8$ 
divergence. When $M=M^*$, the solution has a conical singularity where 
$g_{tt}$ is non-vanishing.  Thus, for spherically-symmetric solutions with 
$k=1$, we have the following classifications:
\begin{itemize}

\item $Q>\ft{2}{b}$:
\begin{itemize}
\item $M>M^*$: Schwarzschild-like black hole with space-like $1/r^6$ singularity.
\item $M=M^*$: Black hole with space-like $1/r^4$ conical singularity.
\item $M_{\rm ext}<M<M^*$: Black hole with time-like $1/r^6$ singularity, 
with outer and inner horizons.
\item $M=M_{\rm ext}$: Extremal black hole with time-like $1/r^6$ singularity.
\item $M<M_{\rm ext}$: Naked time-like $1/r^6$ singularity.
\end{itemize}

\item $Q<\ft{2}{b}$:
\begin{itemize}

\item $M>M^*$: Schwarzschild-like black hole with space-like $1/r^6$ singularity.
\item $M=M^*$: Naked time-like $1/r^4$ singularity.
\item $M<M^*$: Naked time-like $1/r^6$ singularity.
\end{itemize}

\item $Q=\ft{2}{b}$:
\begin{itemize}
\item $M>M^*$: Schwarzschild-like black hole with space-like $1/r^6$ singularity.
\item $M=M^*$: A null singularity where the horizon and curvature 
singularity coincide.
\item $M<M^*$: Naked time-like $1/r^6$ singularity.
\end{itemize}

\end{itemize}
It is worth noting that extremal black hole arises only for $Q>2/b$.

As mentioned earlier, the matter field $A$ is singularity free at $r=0$, one 
would then expect that there exists a parameter like $M=M^*$ such that the 
spacetime solution is free from singularity.  However, there is a singularity 
for the general solutions.  To understand this phenomenon, we note from (\ref{eq9}) 
that the matter energy-momentum tensor is
\be
T^{\mu\nu}_{\rm mat}=-\frac{b^2}{2} \frac{\sqrt{-h}}{\sqrt{-g}} \,\left(h^{-1} 
\right)^{(\mu\nu)}\,.
\ee
It follows that even if $h_{\mu\nu}$ is non-singular and non-vanishing at $r=0$, 
the matter energy-momentum tensor diverges at $r=0$ since $\sqrt{-g}$ vanishes 
there.  The singularity however becomes much milder, and the solution with $Q<2/b$ 
and $M=M^*$ may be viewed as a quasi-soliton.


\section{Generalization to higher dimensions}\label{sec4}

In previous sections, we studied the dyonic black hole solutions in the 
four-dimensional EBI theory, and obtained the first law of thermodynamics for these 
black holes. Now in this section, we will generalize these results to arbitrary 
even dimensions $D=2+2n$.


\subsection{Local solutions}\label{sec4A}

The general ansatz for AdS planar black holes in $D = 2 + 2 n$ dimensions 
is given by
\begin{equation}\begin{split}
ds^2 &= - f(r) dt^2 + \frac{dr^2}{f(r)} + r^2\left(dx_1^2 + dx_2^2 + \cdots 
+ dx_{2n-1}^2 + dx_{2n}^2\right) , \\
F &= \phi^\prime(r)dr \wedge dt + p\left(dx_1\wedge dx_2 +\cdots +dx_{2n-1}
\wedge dx_{2n}\right) \,.\label{eq33}
\end{split}\end{equation}
The equations of motion of $A_\nu$ imply that
\be{}
\phi(r) = \int_r^\infty\frac{qdr}{\sqrt{(r^4 + \frac{p^2}{b^2})^n + 
		\frac{q^2}{b^2}}} \,.\label{eq34}
\ee
It reduces to the previous $D=4$ case when $n=1$. The Einstein equations imply that
\be{}
(r^{2n-1} f)'  = -\fft{\Lambda_0}{n} r^{2n} - \frac{b^2}{2n}
\sqrt{\left(r^4 + \frac{p^2}{b^2}\right)^n + \frac{q^2}{b^2}} \,.\label{eq35}
\ee
Note that $\Lambda_0=\Lambda - \ft12b^2$ and hence there is a smooth 
$b\rightarrow \infty$ limit.  However, the limit $b\rightarrow 0$ is singular
 for $n\ge 3$.  

Thus we see that the solution to the metric function can be expressed in terms of
 a quadrature.  In order to read off the thermodynamical quantities, we would like 
 to write the solution in terms of a well-defined quadrature as a definite 
 integration .  To do so, we may define a function $U_n(r)$, which is convergent 
 at $r=0$, such that $U_n(r)-\sqrt{(r^4 + \ft{p^2}{b^2})^n + \ft{q^2}{b^2}}$ has 
 a falloff that is faster than $1/r$.  This choice is not unique, and we may choose 
 $U_n=(r^4 + p^2/b^2)^{\fft{n}2}$.  Making use of the identity
\be{}
\int_0^r d\tilde r\, \left(\tilde r^{2n}-\Big(\tilde r ^4 + 
\fft{p^2}{b^2}\Big)^{\fft{n}2}\right) = \fft{r^{2n+1}}{2n+1} \left(1 - 
{}_2F_1\left[-\fft14-\fft{n}2,-\fft{n}2; \fft34-\fft{n}2; -\fft{p^2}{b^2 
	r^4}\right]\right) \, ,
\ee
we find that the function $f$ can now be expressed as
\bea
f(r)
&=&  -\frac{\mu}{r^{2n-1}} -\frac{\Lambda_0}{n (2 n + 1)} r^2 - \fft{b^2r^2}{2n(2n+1)}
 {}_2F_1\left[-\fft14-\fft{n}2,-\fft{n}2; \fft34-\fft{n}2; -\fft{p^2}{b^2 r^4}\right]\cr
&&+ \frac{b^2}{2 n r^{2n-1}} \int^r_{\rm \infty} d\tilde{r} \left\{ \Big(\tilde r^4 + \fft{p^2}{b^2}\Big)^{\fft{n}2}-\sqrt{\left(\tilde{r}^4 + \frac{p^2}{b^2}\right)^n + 
	\frac{q^2}{b^2}}\right\}\,.\label{eq37}
\eea


\subsection{Thermodynamics}\label{sec4B}

Now we study the thermodynamics of the dyonic AdS planar black holes in $D = 2+2n$ 
dimensions, constructed in the previous subsection.  In the large-$r$ expansion, 
the term associated with graviton condensation has the falloff of $1/r^{2n-1}$.  
It follows from (\ref{eq37}) that its coefficient is $-\mu$, with no other terms 
giving any further contribution.  Although there are slower falloffs due to the 
presence of the magnetic charges, one can nevertheless, following from the Wald 
formalism, define a ``gravitional mass'' associated with only the condensation of 
the graviton modes \cite{Liu:2015tqa}.  It is given by
\be{}
M = \frac{n\omega_2^n}{8\pi}\mu \,,\label{eq39}
\ee
where
\begin{equation}\begin{split}
\mu &= -\frac{\Lambda_0 {r_+}^{2n+1}}{n (2 n + 1)} -\fft{b^2r_+^{2n+1}}{2n(2n+1)}  
 {}_2F_1\left[-\fft14-\fft{n}2,-\fft{n}2; \fft34-\fft{n}2; -\fft{p^2}{b^2 r_+^4}\right]  \\
&\quad + \frac{b^2}{2 n} \int^{r_+}_{\rm \infty} d{r} \left\{ \Big(r^4 + 
\fft{p^2}{b^2}\Big)^{\fft{n}2}-\sqrt{\left({r}^4 + \frac{p^2}{b^2}\right)^n + 
	\frac{q^2}{b^2}}\right\}\,. \label{eq39a}
\end{split}\end{equation}
Here, for simplicity, we assume that $\int dx_1 dx_2=\int dx_3 dx_4=\cdots=
\int dx_{2n-1} dx_{2n}\equiv \omega_2$. The relation between the mass and the 
horizon radius $r_+$ can be determined by $f(r_+)=0$.  We can now treat $(q,p,r_+)$ 
as independent parameters of the solution.  In terms of these parameters, the 
temperature and entropy are given by
\begin{align}
T & = \frac{f^\prime(r_+)}{4\pi} = -\frac{\Lambda_0}{4 n\pi} r_+ - 
\frac{b^2 r_+^{1-2n}}{8 n\pi} \sqrt{\left(r_+^4 + \frac{p^2}{b^2}\right)^n + 
	\frac{q^2}{b^2}} \,, \label{eq40} \\
S &= \frac{\cal A}{4} = \frac{\omega_2^n}{4} r_+^{2n} \,.\label{eq41}
\end{align}
The electric and magnetic charges  are given by
\be{}
Q_e = \frac{\omega_2^n}{16\pi}\sqrt{-h}(h^{-1})^{[tr]}|_{r\rightarrow\infty} = 
\frac{q}{16\pi}\omega_2^n \,, \qquad Q_m = \frac{n\omega_2}{16\pi}\int F = 
\frac{n p}{16\pi}\omega_2 \,,\label{eq42}
\ee
It follows from (\ref{eq34}) that the electric potential is given by
\be{}
\Phi_{e} =\int^\infty_{r_+} \frac{q dr}{\sqrt{\left(r^4 + \frac{p^2}{b^2} \right)^n 
		+\frac{q^2}{b^2}}} \,.\label{eq43}
\ee
It is easy to very that $\Phi_e=\partial M/\partial Q_e$.  By assuming the 
differential first law of black hole thermodynamics (\ref{eq27}) is still hold, we 
can obtain the magnetic potential
\begin{equation}\begin{split}
\Phi_{m} &=  \frac{\partial M}{\partial Q_m} = 2\omega_2^{n-1}\Biggm[
\frac{b^2r_+^{2n+1}}{4np}\left(\left[1+\frac{p^2}{b^2r_+^4}
\right]^{\frac{n}{2}} -{}_2F_1\left[-\frac{1}{4}-\frac{n}{2}, -\frac{n}{2}; 
\frac{3}{4}-\frac{n}{2}; -\frac{p^2}{b^2r_+^4}\right]\right)  \\
&\quad +\frac{p}{2} \int_\infty^{r_+} \left(r^4 + \frac{p^2}{b^2}\right)^{\frac{n}{2}-1}
\left(1-\sqrt{\frac{\left(r^4 + \frac{p^2}{b^2} \right)^n}{\left(r^4 + \frac{p^2}{b^2} 
		\right)^n+\frac{q^2}{b^2}}}\right)dr \Biggm] \,. \label{eq44}
\end{split}\end{equation}
The generalized ``pressure" ${\cal P}_{\Lambda_0} = -\Lambda_0/(8\pi)$, its conjugate 
quantity ${\cal V}$
\be{}
{\cal V} = \frac{\partial M}{\partial {\cal P}_{\Lambda_0} } = \frac{\omega_2^n}{2n+1} 
r_+^{2n+1} \,.\label{eq45}
\ee
The conjugate term of ${\cal P}_b =-b^2/(16\pi)$ is given by
\begin{equation}\begin{split}
{\cal V}_b &= -2n\omega_2^n \Biggm[-\frac{r_+^{2n+1}}{8n}\left(1+\frac{p^2}{b^2r_+^4}
\right)^{\frac{n}{2}} +\frac{2n-3}{8n(2n+1)} r_+^{2n+1} {}_2F_1\left[-\frac{1}{4}-
\frac{n}{2}, -\frac{n}{2}; \frac{3}{4}-\frac{n}{2}; -\frac{p^2}{b^2r_+^4}\right]\\
&+\quad \frac{1}{2n} \int_\infty^{r_+}  \left\{ \Big(r^4 + \fft{p^2}{b^2}\Big)^{\fft{n}2}
-\sqrt{\left({r}^4 + \frac{p^2}{b^2}\right)^n + \frac{q^2}{b^2}}\right\} dr +
\frac{q^2}{4nb^2}\int_\infty^{r_+} \frac{dr}{\sqrt{\left({r}^4 + \frac{p^2}{b^2}\right)^n 
		+ \frac{q^2}{b^2}}}\\
&\quad -\frac{p^2}{4b^2} \int_\infty^{r_+} \left(r^4 + \frac{p^2}{b^2}\right)^{\frac{n}{2}
	-1}\left(1-\sqrt{\frac{\left(r^4 + \frac{p^2}{b^2} \right)^n}{\left(r^4 + \frac{p^2}{b^2}
		 \right)^n+\frac{q^2}{b^2}}}\right)dr \Biggm] \,,\label{eq46}
\end{split}\end{equation}
The extended differential first law of black hole thermodynamics is given by
\be{}
d M = T d S + \Phi_e d Q_e + \Phi_m d Q_m +{\cal V} d {\cal P}_{\Lambda_0} +{\cal V}_b
 d{\cal P}_b\,.\label{eq47}
\ee
The above first law can also be expressed as
\be{}
d M = T d S + \Phi_e d Q_e + \Phi_m d Q_m +{\cal V} d {\cal P}_{\Lambda} +\frac{b}{8\pi}
({\cal V}-{\cal V}_b) d b\,.\label{eq47a}
\ee
The Smarr formula is given by
\be{}
M = \frac{2n}{2n-1}T S - \frac{2}{2n-1}{\cal V P} + \Phi_e Q_e + \frac{1}{2n-1} \Phi_m
 Q_m - \frac{2}{2n-1}{\cal V}_b\, {\cal P}_b \,.\label{eq48}
\ee
The generalized Smarr formula is given by
\be{}
M = \frac{2n}{2n+1}(T S + \Phi_e Q_e) + \frac{2}{2n+1} \Phi_m Q_m \,.\label{eq48a}
\ee
In order to show the above identities, we need to use
\begin{equation}\begin{split}
&\quad \int_\infty^{r_+}  \left\{ \Big(r^4 + \fft{p^2}{b^2}\Big)^{\fft{n}2}-
\sqrt{\left({r}^4 + \frac{p^2}{b^2}\right)^n + \frac{q^2}{b^2}}\right\} dr \\
&= \frac{r_+}{2n+1}\left[\left(r_+^4+\frac{p^2}{b^2}\right)^{\frac{n}{2}} - \sqrt{\left(r_+^4+\frac{p^2}{b^2}\right)^{n}+\frac{q^2}{b^2}} \right] - 
\frac{2n q^2}{(2n+1)b^2}\int_\infty^{r_+} \frac{dr}{\sqrt{\left({r}^4 + 
		\frac{p^2}{b^2}\right)^n + \frac{q^2}{b^2}}}\\
&\quad+\frac{2np^2}{(2n+1)b^2} \int_\infty^{r_+} \left(r^4 + 
\frac{p^2}{b^2}\right)^{\frac{n}{2}-1}\left(1-\sqrt{\frac{\left(r^4 + \frac{p^2}{b^2}
		 \right)^n}{\left(r^4 + \frac{p^2}{b^2} \right)^n+\frac{q^2}{b^2}}}\right)dr \,.
\end{split}\end{equation}
Note that the definite integrations in all the above equations are well-defined with 
no divergence. It is remarkable that although the general solution is given up to a 
well-defined quadrature, the first law of thermodynamics, and Smarr relations can 
nevertheless be fully established.

\subsection{Some explicit examples}\label{sec4C}

We obtained the general dyonic AdS planar black holes, up to a quadrature.  
Here we present some explicit examples where the quadrature can be integrated 
in terms of some special functions.


\subsubsection{Pure electric solutions}\label{sec4c1}

In this case, we set $p=0$, and we find
\begin{equation}\begin{split}
f(r) &= -\frac{\mu}{r^{2n-1}} -\frac{ \Lambda_0r^2 }{n(2n+1)}   -
\frac{b^2 r^{2}}{2n(2n+1)} {}_2F_1\left[-\frac{1}{2}, -\frac{1}{2}-\frac{1}{4n}; 
\frac{1}{2}-\frac{1}{4n}; -\frac{q^2}{b^2r^{4n}} \right] \,. \label{eq49}
\end{split}\end{equation}
Although our ansatz is for even $D = 2n +2$ dimensions, the above solution is applicable 
for odd dimensions as well so we rewrite it in terms of $D$:
\begin{equation}\begin{split}
f(r) &= -\frac{\mu}{r^{D-3}} -\frac{2\Lambda_0r^2 }{(D-2)(D-1)}\\
  &+ \frac{b^2 r^{2}}{(D-2)(D-1)} {}_2F_1\left[-\frac{1}{2}, -\frac{D-1}{2(D-2)}; 
  \frac{D-3}{2(D-2)}; -\frac{q^2}{b^2r^{2D-4}} \right] \,.\label{eq50}
\end{split}\end{equation}
Furthermore, we can add a topological parameter $k$ to $f$ so that $f\rightarrow f + k$. 
The solution becomes that for general topologies and was obtained in \cite{Cai:2004eh}.


\subsubsection{Pure magnetic solutions}\label{sec4c2}

In this case, we set $q=0$, and we find
\be{}
f(r) = -\frac{\mu}{r^{2n-1}} -\frac{\Lambda_0}{n (2 n + 1)} r^2 - \fft{b^2r^2}{2n(2n+1)}
 {}_2F_1\left[-\fft14-\fft{n}2,-\fft{n}2; \fft34-\fft{n}2; -\fft{p^2}{b^2 r^4}\right] 
 \,.\label{eq51}
\ee
Note that when $n=1$, corresponding to four dimensions, the two solutions (\ref{eq51}) 
and (\ref{eq50}) takes the same form with $q\leftrightarrow p$, indicating electric and 
magnetic duality.

Note also that when $n = 2m$ is even, corresponding to $D = 4 m +2 = 6, 10, \dots$ 
dimensions, the hypergeometric function in (\ref{eq51}) solution reduces to some 
polynomial functions. Here are some low-lying examples in 6, 10, 14 respective dimensions:
\begin{equation}\begin{split}
n &= 2 : f(r) = -\frac{\Lambda}{10}r^2 -\frac{\mu}{r^3} - \frac{p^2}{4r^2} \,; \\
n &= 4 : f(r) = -\frac{\Lambda}{36}r^2 -\frac{\mu}{r^7} - \frac{p^2}{20r^2} - 
\frac{p^4}{8 b^2 r^6} \,; \\
n &= 6 : f(r) = -\frac{\Lambda}{78}r^2 -\frac{\mu}{r^{11}} - \frac{p^2}{36r^2} - 
\frac{p^4}{20 b^2 r^6} - \frac{p^6}{12 b^4 r^{10}} \,. \label{eq52}\\
\end{split}\end{equation}
Note that when $n=2$, corresponding to $D=6$, the metric is independent of $b$, which 
implies that the energy-momentum tensor for the Born-Infeld model is the same as that 
of the Maxwell theory.

Here we present the thermodynamical properties of pure magnetic AdS planar black holes
\begin{align}
\mu &= -\frac{\Lambda_0 {r_+}^{2n+1}}{n (2 n + 1)} -\fft{b^2r_+^{2n+1}}{2n(2n+1)}  
 {}_2F_1\left[-\fft14-\fft{n}2,-\fft{n}2; \fft34-\fft{n}2; -\fft{p^2}{b^2 r_+^4}\right] \,,  \\
T & = -\frac{\Lambda_0}{4 n\pi} r_+ - \frac{b^2 r_+^{1-2n}}{8 n\pi} \left(r_+^4 + 
\frac{p^2}{b^2}\right)^{\frac{n}{2}} \,,  \qquad S =  \frac{\omega_2^n}{4} r_+^{2n} 
\,,\label{eq41a} \qquad Q_m =  \frac{n p}{16\pi}\omega_2 \,, \\
\Phi_{m} &=  2\omega_2^{n-1}\Biggm[\frac{b^2r_+^{2n+1}}{4np}\left(\left[1+\frac{p^2}{b^2r_+^4}
\right]^{\frac{n}{2}} -{}_2F_1\left[-\frac{1}{4}-\frac{n}{2}, -\frac{n}{2}; \frac{3}{4}-\frac{n}{2}; -\frac{p^2}{b^2r_+^4}\right]\right) \Biggm] \,, \\
\begin{split}
{\cal V}_b &= -2n\omega_2^n \Biggm(-\frac{r_+^{2n+1}}{8n}\left(1+\frac{p^2}{b^2r_+^4}
\right)^{\frac{n}{2}} \\
&\quad +\frac{2n-3}{8n(2n+1)} r_+^{2n+1} {}_2F_1\left[-\frac{1}{4}-\frac{n}{2}, -
\frac{n}{2}; \frac{3}{4}-\frac{n}{2}; -\frac{p^2}{b^2r_+^4}\right] \Biggm) \,.\label{eq46a}
\end{split}
\end{align}


\subsubsection{Dyonic solutions}\label{sec4c3}

The quadrature cannot be integrated for general $n$ in terms of a special function, except 
for $n=1$ and $n=2$. The $n=1$ example was discussed earlier. Now let us consider $n=2$. 
The function $f(r)$ is given by
\begin{equation}\begin{split}
f(r) &= -\frac{\Lambda_0}{10}r^2 - \frac{\tilde{\mu}}{r^3} \\
&\quad -\frac{1}{4r^2} \sqrt{p^4+b^2q^2}\,  F_1\left[\frac{1}{4}; -\frac{1}{2}, 
-\frac{1}{2}; \frac{5}{4}; -\frac{b^2r^4}{\sqrt{-b^2q^2}+p^2}; \frac{b^2r^4}{\sqrt{-b^2q^2}-p^2}
 \right] \,,\label{eq53}
\end{split}\end{equation}
where $F_1$ is the Appell hypergeometric function. This form of the solution is not 
convenient for extracting the asymptotic infinite behavior. Another equivalent form of 
the solution is given by
\begin{equation}\begin{split}
f(r) &= -\frac{\Lambda_0}{10}r^2 - \frac{\mu}{r^3} - \frac{4p^2+b^2r^4}{20r^6} 
\sqrt{\frac{q^2}{b^2} + \left(r^4 + \frac{p^2}{b^2} \right)^2} \\
&\quad +\frac{3p^4+b^2q^2}{15b^2r^6} F_1\left[\frac{3}{4}; \frac{1}{2}, \frac{1}{2}; 
\frac{7}{4}; \frac{\sqrt{-b^2q^2} -p^2}{b^2r^4}, -\frac{\sqrt{-b^2q^2} -p^2}{b^2r^4}  \right] \\
&\quad +\frac{3p^2(p^4+b^2q^2)}{35b^4r^{10}} F_1\left[\frac{7}{4}; \frac{1}{2}, \frac{1}{2}; 
\frac{11}{4}; \frac{\sqrt{-b^2q^2} -p^2}{b^2r^4}, -\frac{\sqrt{-b^2q^2} -p^2}{b^2r^4} 
 \right] \,.\label{eq54}
\end{split}\end{equation}
The large-$r$ expansion of $f(r)$ is given by
\begin{equation}\begin{split}
f(r) &= -\frac{\Lambda}{10}r^2 -\frac{p^2}{4r^2} - \frac{{\mu}}{r^3} +\frac{q^2}{24r^6} 
-\frac{p^2q^2}{56b^2r^{10}} - \frac{q^2(b^2q^2 -4p^4)}{352b^4r^{14}} \\
&\quad +\frac{p^2q^2(3b^2q^2 -4p^4)}{480b^6r^{18}} +\cdots \,.\label{eq55}
\end{split}\end{equation}
It is then clear that the parameter $\mu$ is related to the gravitional mass.  Since this 
solution is a special case of the general solutions, we shall not discuss its thermodynamics 
further.


\subsubsection{A more general topology}\label{sec4c4}

We may consider more general ansatz with the following general topologies in $D=2n+2$ dimensions
\begin{align}
ds^2 &= -f(r)dt^2 +\frac{dr^2}{f(r)} + r^2\Sigma_{i=1}^n d\Omega_{i,k}^2 \,,\\
F &= \phi^\prime(r) dr\wedge dt + p\,\Sigma_{i=1}^n dx_i\wedge dy_i \,,
\end{align}
where
\be{}
d\Omega_{i,k}^2 = \frac{dx_i^2}{1-kx_i^2} +(1-kx_i^2)dy_i^2 \,.
\ee
The $\phi(r)$ is given again the same as (\ref{eq34}) and $f(r)$ is given by
\begin{equation}\begin{split}
f(r) &= \frac{k}{2n-1} -\frac{\mu}{r^{2n-1}} -\frac{\Lambda_0}{n (2 n + 1)} r^2 - 
\fft{b^2r^2}{2n(2n+1)}
 {}_2F_1\left[-\fft14-\fft{n}2,-\fft{n}2; \fft34-\fft{n}2; -\fft{p^2}{b^2 r^4}\right] \\
&\quad + \frac{b^2}{2 n r^{2n-1}} \int^r_{\rm \infty} d\tilde{r} \left\{ \Big(\tilde r^4 + \fft{p^2}{b^2}\Big)^{\fft{n}2}-\sqrt{\left(\tilde{r}^4 + \frac{p^2}{b^2}\right)^n + 
	\frac{q^2}{b^2}}\right\} \,.
 \end{split}\end{equation}
It reduces to the previous result (\ref{eq37}) for $k =0$.  The horizon topology now 
becomes ${\cal M}_2\times {\cal M}_2\times \cdots\times {\cal M}_2$, where ${\cal M}_2$ 
can be sphere, torus or hyperbolic 2-space.


\section{Conclusion}\label{sec5}

In this paper, we studied the EBI theory and derived the equations of motion that is 
valid in all dimensions and for all charge configurations.  By contrast, the Lagrangian 
(\ref{eq3}) considered in many previous works has limited application in higher 
dimensions. We then constructed the dyonic AdS black holes in four dimensions with a 
general topology. We analyzed the global structure and obtained the first law of 
thermodynamics.  We classified the singularity structure of these solutions.  We then 
constructed the dyonic AdS black holes in general even dimensions, where the solutions 
carry both the electric charge and also the magnetic fluxes along the planar space.  The 
general solutions were given up to a quadrature; nevertheless, we show that the first 
law of black hole thermodynamics can be established.  We also give many special examples 
where the quadrature can be integrated in terms of special functions.  These solutions 
provide new gravity duals to study the AdS/CFT correspondence.


\section*{ACKNOWLEDGEMENTS}
S.L.L.~is grateful to Shuang-Qing Wu for useful discussions. He also thanks Xing-Hui 
Feng for kind help and discussions. S.L.L. and H.W. are supported in part by 
NSFC under Grants No.~11575022 and No.~11175016. H.L.~is supported in part by 
NSFC grants NO. 11175269, No.~11475024 and No.~11235003.

\renewcommand{\baselinestretch}{1.0}

\end{document}